\documentclass[a4 paper, 12 pt] {article}

\begin{document}
\title{\bf{Quark-antiquark potentials from a scalar field in SU(2) YM}}
\author{M. \'{S}lusarczyk\thanks{mslus@alphas.if.uj.edu.pl} \,
and A. Wereszczy\'{n}ski\thanks{wereszcz@alphas.if.uj.edu.pl}
       \\ Institute of Physics,
       \\ Jagellonian University, Reymonta 4, Krakow, Poland}
\maketitle
\begin{abstract}
   A generalized Dick model with a potential term is discussed. The
   solution originating from a static, pointlike, color source
   is found to have a confining part. The comparison with a
   wide spectrum of phenomenological quark-antiquark potentials is
   presented.
\end{abstract}

%%%%%%%%%%%%%%%%%%%%%%%%%%%%%%%%%%%%%%%%%%%%%%%%%%%%%%%%%%%%%%%%%%%%%
\section{\bf{The model}}
%%%%%%%%%%%%%%%%%%%%%%%%%%%%%%%%%%%%%%%%%%%%%%%%%%%%%%%%%%%%%%%%%%%%%
It is believed that the nonrelativistic potential model can quite
well describe the physics of heavy quarks. It is possible to
obtain the whole spectrum of the masses of the quark and antiquark
pairs in the quarkonium system, using only potential $U(r)$. Here
$r$ denotes a distance between quarks. Unfortunately, there are
many different forms of the potential in the literature (see eg.
\cite{cornel}, \cite{potential}) nevertheless up to now the final
form has not been fixed. However, it was shown \cite{motyka} that
the most probable potential, in the bottomonia region, is:
\begin{equation}
U_{MZ}(r)=C_1 \left(\sqrt{r} -\frac{C_2}{r} \right),
\label{potkacper}
\end{equation}
where $C_1\simeq 0.71$ Gev$^{\frac{1}{2}}$ and $C_2 \simeq 0.46$
Gev$^{\frac{3}{2}}$. In the present paper we construct a Lorentz
invariant, effective action which provides this potential.
\newline
The confining part of the potential, i.e. the part which is
divergent in spatial infinity, can be obtained from the $SU(2)$
Yang-Mills theory coupled to the scalar field \cite{dick1},
\cite{dick2}, \cite{chabab}. In particular, in the generalized
Dick model \cite{my}, there is a sector where confining potential
has the same form as that given by (\ref{potkacper}). The main
defect of the generalized Dick model (from confinement point of
view) is the simultaneous existence of the finite energy solutions
of the Coulomb problem. In order to preserve only the confining
sector one has to add an additional potential term for the scalar
field. This potential should have a unique minimum at $\phi =0$
\cite{my}. On the other hand, the Motyka-Zalewski potential
(\ref{potkacper}) contains also the well-known, Coulomb part.
Because of that our effective action should, in the limit of a
strong scalar field, reduce to the simple YM theory. The effective
action discussed below can also be regarded as a version of the
$SU(2)$ color dielectric model with the special choice of the
color dielectric function and scalar potential (see eg.
\cite{pirner1}).
\newline
Let us consider a model which satisfies the conditions mentioned
above:
\begin{equation}
S = \int d^{4}x \left[ -\frac{1}{4} \frac{
(\frac{\phi}{\Lambda})^{8\delta}}{1+(\frac{\phi}{\Lambda})^{8\delta}}
 F^{a}_{\mu
\nu} F^{a \mu \nu} + \frac{1}{2} \partial_{\mu} \phi
\partial^{\mu} \phi  - \alpha \phi^4
\left( \frac{\phi}{\Lambda} \right)^{8\delta} \right],
\label{action}
\end{equation}
where $\delta >0$, $\Lambda $ is a dimensional whereas $\alpha $ a
dimensionless  constant and $F_{\mu \nu}^a$ is defined in the
standard manner.We would like to stress that the potential for the
scalar field has been chosen to give an analytic solution of the
Coulomb problem.
\newline
The field equations for (\ref{action}) have the following form:
\begin{equation}
 D_{\mu} \left( \frac{
(\frac{\phi}{\Lambda})^{8\delta}}{1+(\frac{\phi}{\Lambda})^{8\delta}}
F^{a \mu \nu} \right)= j^{a \nu},
\label{f_eq_1}
\end{equation}
\begin{equation}
\partial_{\mu} \partial^{\mu} \phi = -2\delta F^{a}_{\mu \nu} F^{a \mu \nu}
\frac{ \phi^{8 \delta - 1}}
{\Lambda^{8\delta }(1+\frac{ \phi^{8 \delta}}{\Lambda^{8\delta }})^2}
-\alpha (4+8\delta ) \frac{\phi^{8\delta +3}}{\Lambda^{8\delta}} ,
\label{f_eq_2}
\end{equation}
where $j^{a \mu}$ is the external color current.
%%%%%%%%%%%%%%%%%%%%%%%%%%%%%%%%%%%%%%%%%%%%%%%%%%%%%%%%%%%%%%%%%%%%%
\section{Solutions}
%%%%%%%%%%%%%%%%%%%%%%%%%%%%%%%%%%%%%%%%%%%%%%%%%%%%%%%%%%%%%%%%%%%%%
Let us consider a static, pointlike color source:
\begin{equation}
j^{a \mu} = 4\pi q \delta(r) \delta^{a3} \delta ^{\mu 0},
\label{source}
\end{equation}
Without loss of generality the Abelian source can be taken. One
can consider a non-Abelian source, for example: $j^{a \mu} = 4\pi
q \delta(r) C^a \delta ^{\mu 0}$, where $C^a$ is the expectation
value of the su($N_c$) generator for a normalized spinor in the
color space \cite{dick1}. However, on account of the fact that the
results for these two cases are very similar we will analyze only
the Abelian source. The pertinent equations of motion read:
\begin{equation}
\left[ r^2 \frac{
(\frac{\phi}{\Lambda})^{8\delta}}{1+(\frac{\phi}{\Lambda})^{8\delta}}
E \right]' =4\pi q \delta(r) \label{culomb1}
\end{equation}
\begin{equation}
\nabla^2_r \phi =-4\delta E^2 \frac{ \phi^{8 \delta - 1}}
{\Lambda^{8\delta }(1+\frac{ \phi^{8 \delta}}{\Lambda^{8\delta
}})^2} + \alpha (4+8\delta ) \frac{\phi^{8\delta
+3}}{\Lambda^{8\delta}},
\end{equation}
where $\vec{E}^a=E \delta^{3a} \hat{r}$ is the electric field
defined in the standard way. These equations posses the following
solutions:
\begin{equation}
\phi (r)=\Lambda A \left( \frac{1}{\Lambda r }
\right)^{\frac{1}{1+4\delta}}
\label{sol1}
\end{equation}
\begin{equation}
\vec{E} (r) = \left[ \frac{q}{r^2} + A^{-8\delta} q\Lambda^2
\left( \frac{1}{\Lambda r} \right)^{\frac{2}{1+4\delta }} \right]
\hat{r}, \label{sol2}
\end{equation}
where $A=\left[ \frac{-4\delta + \sqrt{4\delta (1+4\alpha )
(1+2\delta)(1+4\delta )^4)}}{8\alpha (1+2\delta )(1+4\delta )^2}
\right]^{\frac{1}{2(1+4\delta)}}$. This number is positive for
sufficiently large values of $\alpha$. The energy of the solutions
is divergent, not only in the small $r$ limit but also in the long
range limit, $r \rightarrow \infty$. In that sense the confinement
emerges.
\newline
For $\delta > \frac{1}{4}$ the color-electric potential has the
form:
\begin{equation}
 V(r) =
   -\frac{q}{r}+ \frac{4\delta +1}{4\delta -1}
  A^{-8\delta } q \Lambda^{\frac{4\delta }{4\delta +1}} \cdot
  \, r^{\frac{4\delta -1}{4\delta +1}}.
\label{potential1}
\end{equation}
Let us assume that a color source is a heavy quark. Therefore, the
potential seen by an (anti)quark has the following form:
\begin{equation}
   U(r) =
   -\frac{q^2}{r}+ \frac{4\delta +1}{4\delta -1}
  A^{-8\delta } q^2 \Lambda^{\frac{4\delta }{4\delta +1}} \cdot
  \, r^{\frac{4\delta -1}{4\delta +1}}.
\label{potential2}
\end{equation}
The similar calculation can be done for the case  $\delta =\frac{1}{4}$.
The result is:
\begin{equation}
 U(r)= -\frac{q^2}{r} +\Lambda A^{-8\delta} q^2 \ln \Lambda r.
\label{potln}
\end{equation}
For $\delta <\frac{1}{4}$ the potential does not show
confinement-like behaviour.
\newline
There are three general conditions which must be satisfied by a
static potential. Namely, it cannot rise faster than linearly as a
function of the distance $r$ for $r \rightarrow \infty $
\cite{seiler}, it has to be a monotonically rising function of $r$
and $U''(r) \leq 0$ \cite{bachas}. Unfortunately, our potential
(\ref{potential2}) satisfies these conditions for all $\delta
>\frac{1}{4}$ and we cannot use them to constrain the parameter
$\delta $.
\newline
\newline
Potential term in (\ref{action}) not only excludes single charge
states from the physical sector of the theory (i.e. there are no
finite energy solutions of the Coulomb problem) but also removes
magnetic monopoles. Let us rewrite the equations of motion using
the well-known magnetic monopole Ansatz for the gauge field:
\begin{equation}
A_i^a =\epsilon_{aik} \frac{x^k}{r^2} (g(r)-1), \;\;\; A_0^a=0,
\label{ansatz}
\end{equation}
where $g(r)$ is a function of the radial coordinate only. We get:
\begin{equation}
\left[ \frac{
(\frac{\phi}{\Lambda})^{8\delta}}{1+(\frac{\phi}{\Lambda})^{8\delta}}
g' \right]' + \frac{
(\frac{\phi}{\Lambda})^{8\delta}}{1+(\frac{\phi}{\Lambda})^{8\delta}}\frac{g}{r^{2}}
\left( 1 - g^{2} \right) = 0, \label{f_eq_5}
\end{equation}
\begin{equation}
\frac{1}{r^{2}} \left( r^{2} \phi' \right)' = 4 \delta \frac{
\phi^{8 \delta - 1}} {\Lambda^{8\delta }(1+\frac{ \phi^{8
\delta}}{\Lambda^{8\delta }})^2} \left[ \frac{2g'^{2}}{r^{2}} +
\frac{(g^{2} - 1)^{2}}{r^{4}} \right]  +\alpha (4+8\delta )
\frac{\phi^{8\delta +3}}{\Lambda^{8\delta}}. \label{f_eq_6}
\end{equation}
The above set of equations possesses the unique but trivial finite
energy solution $\phi =0$. The finite energy monopoles observed in
\cite{my} do not appear due to the potential term.
%%%%%%%%%%%%%%%%%%%%%%%%%%%%%%%%%%%%%%%%%%%%%%%%%%%%%%%%%%%%%%%%%%%%%%%
\section{Conclusions}
%%%%%%%%%%%%%%%%%%%%%%%%%%%%%%%%%%%%%%%%%%%%%%%%%%%%%%%%%%%%%%%%%%%%%%%
We can compare the quark-antiquark potential derived from model
(\ref{action}) with the phenomenological confining Motyka-Zalewski
potential. It is immediately seen that they become identical if we
set $\delta = \frac{3}{4}$.
\newline
However, one can fit the model to another phenomenological
potential, which has been successfully applied to calculate
quarkonium energy levels, namely to the Cornell potential
$U_{C}(r)=-\frac{a}{r} +b r$, where $a,b$ are nonnegative
constants \cite{cornel}. Our potential has the same form in the
limit  $\delta \rightarrow \infty$, but it is not feasible to
implement this limit on the lagrangian level. So, strictly
speaking, our model does not supply the linear divergence of the
confining potential but it can be achieved with arbitrary accuracy
by taking sufficiently large values of $\delta$. That is the main
difference between the Dick model and the model presented here.
\newline
Contrary to the model considered in \cite{my}, there exists no
solution which is non-singular in spatial infinity. It was gained
by adding the potential term for the scalar field. There are no
magnetic monopoles, either. It is worth stressing that in the
model the confinement and disappearance of magnetic monopoles
occur simultaneously.
\newline
There are, at least,  two directions in which the present work
might be continued. Firstly, the theoretical restrictions for the
parameter $\delta$ as well as for the particular form of the
potential term in the action (\ref{action}) are needed. Due to the
fact that our action belongs to a wide family of color-dielectric
actions we believe that these problems can be solved using the
lattice color-dielectric methods. In fact much work was done in
the past to derive the lattice color dielectric model from the
lattice QCD (see eg. \cite{pirner2}, \cite{pirner3},
\cite{arodz}). For example in paper \cite{pirner3} the effective
scalar potential for $\delta=\frac{1}{2}$ has been computed. One
can also use the detailed, lattice study of the flux--tube
profile, which was done in the last few years (see eg.
\cite{bali1}, \cite{bali2}, \cite{bali3}) and compare it with the
predictions of our model to fix the $\delta$ parameter. \\
 Secondly, because the finite energy solutions of the
Coulomb problem appear for potentials which have a unique minimum
for $\phi \neq 0$, the model can be used to study
confinement-deconfinement phase transition.

\end{document}